\documentclass[11pt,twoside]{article}
\usepackage{asp2010}

\resetcounters

\begin{document}

\title{Structure of Small Magnetic Elements in the Solar Atmosphere}
\author{Vicente Domingo,$^1$ Judith Palacios,$^1$ Laura A. Balmaceda,$^2$ Santiago~Vargas~Dom\'inguez,$^{3}$ and Iballa Cabello$^{1}$
\affil{$^1$Image Processing Laboratory, Universidad de Valencia, P.O. Box 22085, E-46071, Valencia, Spain}    
\affil{$^2$Inst. de Ciencias Astron\'omicas, de la Tierra y del Espacio - ICATE, CONICET, Av. Espa\~na 1512 Sur, 5400 San Juan, Argentina}
\affil{$^3$ Mullard Space Science Laboratory, University College London, Holmbury St Mary, Dorking, Surrey, RH5 6NT, UK}
}

\begin{abstract} 
High resolution images at different wavelengths, spectrograms and magnetograms,
representing different levels of the solar atmosphere obtained with Hinode have
been combined to study the 3-dimensional structure of the small magnetic
elements in relation to their radiance. A small magnetic element is described as example
of the study.
\end{abstract}

\section{Introduction}
The small magnetic elements have been studied profusely, see for instance the
review paper by \cite{dewijn2009}. However, many essential characteristics
remain contradictory regarding different studies. At the beginning of the 70s,
several studies started to indicate that most of the magnetic flux outside
active regions was contained in sub-resolution kilogauss elements. This
conclusion derived in the ‘thin tube flux’ model \citep{spruit1976,  parker1976}
making these unresolved structures the ‘building block’ or ‘element’ in the
photosphere. However, these features can appear in the shape of thin sheet
 \citep{berger2004}  may be composed of smaller tubes, or tubes that can be
structured as a bunch of slender tubes \citep{sanchezalmeida1997}; they appear
in every environment in the solar photosphere. These structures appear as
facular brightening, G-band bright points, filigree. In this work we focus on G-band
bright points analysis. G-band bright points, GBP, are bright in G-band because inside the small magnetic tube, lateral
heating provokes differential CH dissociation, that makes the opacity lower and
the tube brighter \citep{kiselman2001}. These bright points were also observed
in CN band, their contrast calculated and compared with G-band in filter
bandwidth dependence and MHD modeling by \cite{uitenbroek2006}. In the case of
the GBPs, the detection, size and coverage are usually calculated by
the use of segmentation algorithms, which detect high contrast above a
threshold, separate from the structure where they are embedded (an inter-granular lane
or close to a granule) and merge the structure. A related study, in a joint
campaign with the SST of La Palma, uses Hinode CN filter-grams and Mg~$\textsc{i}$
magnetograms, close to the disk center, on September, 29 2007, and SST G-band
data. MLT4 segmentation code of \cite{bovelet2007}  was applied to CN images, and 
the segmentation code used in \cite{sanchezalmeida2004} was used for G-band SST images, finding that the  area coverage is 0.22
for G-band, 0.26 for CN \citep{balmaceda2008}.

 \section{Observations}
We use two Hinode data sets in three different wavelengths plus the
SOT-SP (Spectro\-polarimeter) data. One data set, obtained in 2008, Aug 6, from 08:02:56 to 09:17:05 UT, consists on
CN ($\lambda$=388.3nm), Ca~$\textsc{ii}$($\lambda$=396.8nm), G-band($\lambda$= 430.56 nm) 
images, SP scans, centered in  $\mu$=0.85 (491'',-14''). The
second data set, from 2008,  Aug 7, from 08:01:24 to 09:06:59 UT, is si\-milar, but also with one Na~$\textsc{i}$ magnetogram 
in the blue wing of the Na~$\textsc{i}$ line  ($\lambda$= 589.6 nm)
and centered in $\mu$=0.86, at coordinates (-8'',485''). The Broadband Filter (BFI) has a FOV of
55''75 $\times$ 55''75, the SP maps 58''$\times$  81'' and the Na~$\textsc{i}$ magnetogram 
31''$\times$ 81''. The reduction procedure is the standard: correction for dark-current and flat-field, and also co-alignment.

\section{Analysis}
\begin{figure}[h!]
 \center
\includegraphics[scale=0.12]{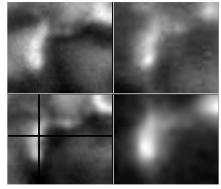}  \hspace{2cm} \\
\caption{Case study of a facula. From upper-left corner in clockwise direction: CN, Ca~$\textsc{ii}$, Na~$\textsc{i}$, G-band.}\label{fig1}
\end{figure}

From the two data sets we have chosen for a detailed analysis four small
elements, and studied their images, contour plots of the contrast of the
images and the magnetic field intensity. We also have drawn the profiles of
their values in two perpendicular directions to visualize their structure. Each
of the sampled elements appear to have a different structure as may be expected
from the turbulent nature of the atmospheric convection. Due to the different
height formation of their radiation we see that the chromospheric Ca~$\textsc{ii}$~H element
images are generally wider than the G and CN bands due to the expansion
of the flux tubes with height. \cite{qu2002} show that the Ca~$\textsc{ii}$ radiation is generated 
in the chromosphere in quiet sun regions. Because of the space limitation we show only one of the
four analyzed elements, corresponding to the Aug 7 dataset. In Fig.~\ref{fig1} from the upper-left corner and in clockwise direction, CN, Ca~$\textsc{ii}$, Na~$\textsc{i}$ magnetogram and G-band images are shown, two straight lines indicate the section through which we have cut the image to derive indicative profiles of the structure. The size of these images is 2''1 $\times$ 1''9.

\begin{figure}[h!]
 \center
\includegraphics[scale = 0.37]{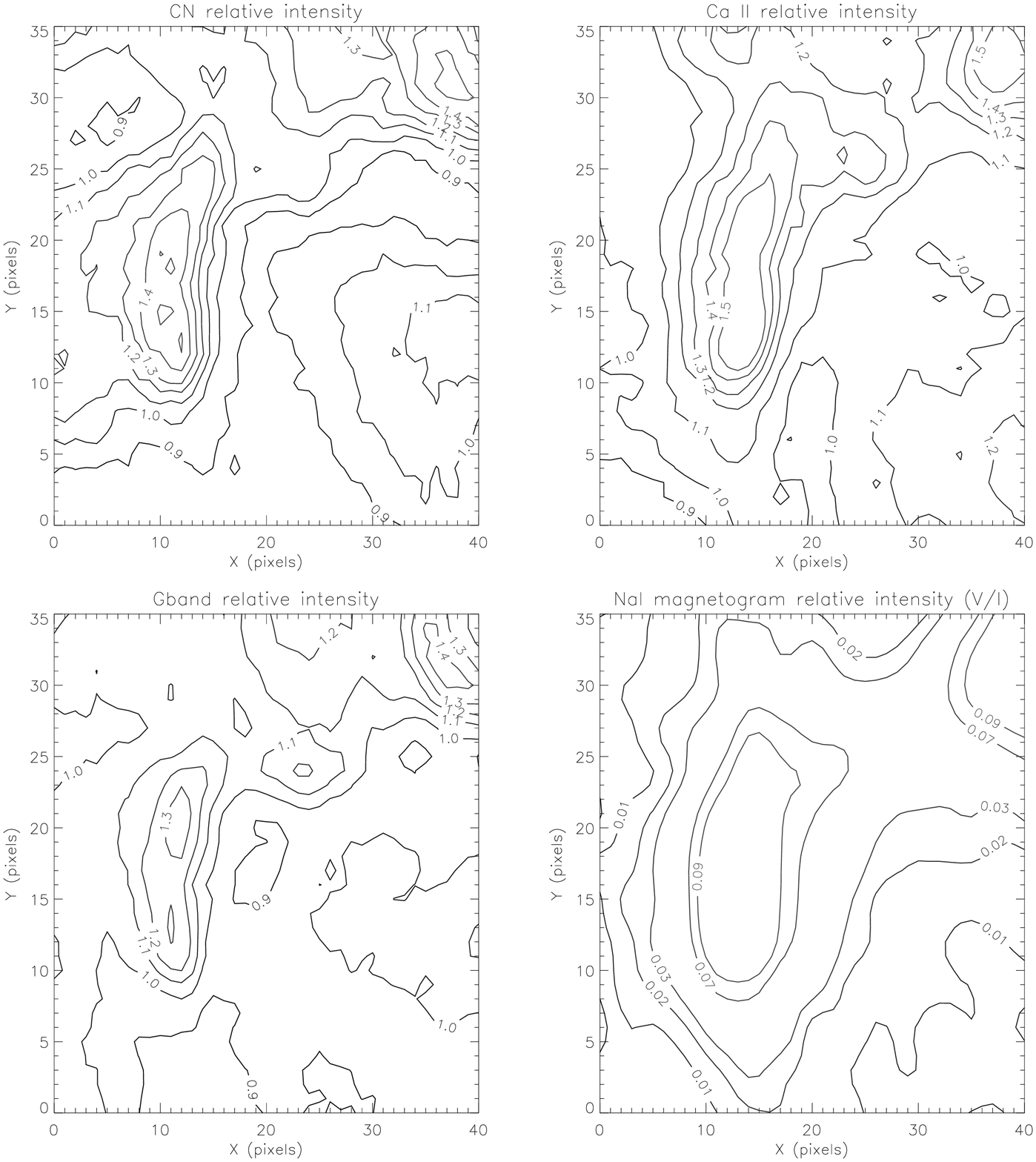}\\
\caption{Relative intensity contours: CN, Ca~$\textsc{ii}$, Na~$\textsc{i}$, G-band}\label{fig2}
\end{figure}

 In Fig.~\ref{fig2} the contours of relative intensity are plotted, in the
same order of display as in Fig.~\ref{fig1}. 
In the contours plots we identify a line of bright points in G-band
and CN band indicating a sort of sheath magnetic field at photospheric level on
the border of a granule, while at chromospheric level, the Na~$\textsc{i}$ magnetogram shows
that the magnetic field bundle is largely expanded, expansion supported by the
extension of the Ca~$\textsc{ii}$ H image. Also, the South-North profile shows bumps in the G
and CN bands, north of the peak in the magnetogram that would correspond to the
well known effect of facular bright points resulting of seeing the outbound side
of the magnetic field tube (the facula is located at $\mu$=0.86).

\begin{figure}
 \center
\includegraphics[scale = 0.35]{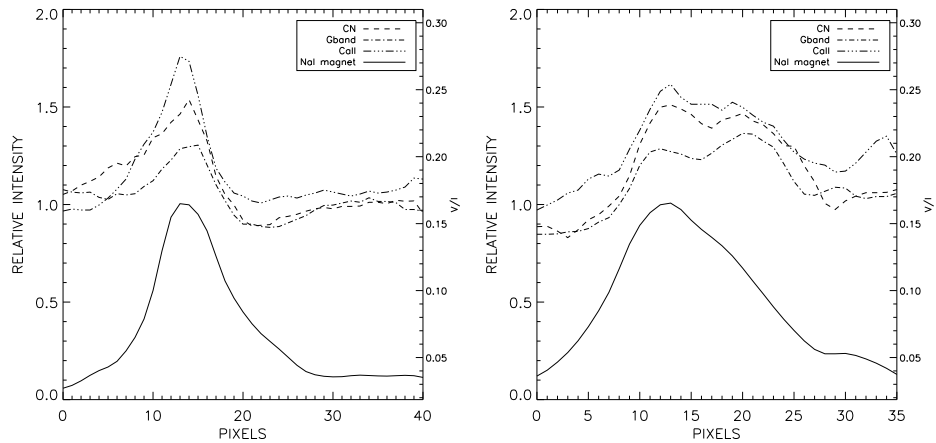}
\caption{West-East (\emph{left}) and North-South (\emph{right})  relative intensity profiles. The scale on the right shows the $V/I$ signal of the Na~$\textsc{i}$ magnetogram}\label{fig3}
\end{figure}

Fig.~\ref{fig3} displays vertical profiles of
the three images and the Na~$\textsc{i}$ magnetogram in two perpendicular planes, one along
the West-East parallel and the other along the South-North meridian, both
indicated in figure 1 by perpendicular lines.  The scale is relative intensity in the left 
side of the plot, and the LOS-magnetogram $V/I$ in the scale of the right side of the plot.
It is interesting to notice that
this bright structure is located at about $\mu$=0.86, near the central North-South
meridian. In the West-East plots, an asymmetry is noticeable from the Western facular 
point to the Eastern inter-granular lane and the adjacent granule.

\section{Discussion and concluding remarks}

Although this is only an example, we can say that the other three studied cases
are also well represented by the expanding magnetic flux tube model with
variations in their detailed structure, due to the diverse local convective
structures, like one case in which there is evidence of the magnetic flux tube
being inclined respect to the local vertical. It is worth reminding that we are studying 
the structure of an extremely dynamic environment. We can see, in general, that the CN band has higher contrast than the G-band in
the small magnetic elements, as was shown by \cite{zakharov2005} in small
magnetic elements in an active region. Due to the different height formation of
their radiation we see that the Ca~$\textsc{ii}$ H element images are generally wider than
the G and CN bands due to the expansion of the magnetic flux bundle with height. 
%
\acknowledgements 
~~\emph{Hinode} is a Japanese mission developed and launched by ISAS/JAXA, with NAOJ as domestic partner and NASA and STFC (UK) as international partners. It is operated by these agencies in co-operation with ESA and NSC (Norway).

\end{document}